\documentclass{llncs}
\usepackage{llncsdoc}
\usepackage{color}

\renewcommand{\symbol}[1]{{\tt #1}}
\newcommand{\attribute}[1]{{\tt '#1'}}
\newcommand{\code}[1]{{\tt #1}}

\title{Creating modular and reusable DSL textual syntax definitions with Grammatic/ANTLR}
\author{Andrey Breslav}
\institute{St. Petersburg State University of Information Technology, Mechanics and Optics
\email{abreslav@gmail.com}
}

\begin{document}
\maketitle
\begin{abstract}
In this paper we present Grammatic -- a tool for textual syntax definition. Grammatic serves as a front-end for parser generators (and other tools) and brings modularity and reuse to their development artifacts. It adapts techniques for separation of concerns from Apsect-Oriented Programming to grammars and uses templates for grammar reuse. 
We illustrate usage of Grammatic by describing a case study: bringing separation of concerns to ANTLR parser generator, which is achieved without a common time- and memory-consuming technique of building an AST to separate semantic actions from a grammar definition.
\end{abstract}

\section{Introduction}\label{Intro}

When adopting a concept of Domain-Specific Languages (DSLs, see \cite{LanguageOriented}) and developing textual syntax for them we are using syntax-related tools extensively.
Thus we need different tools which in most cases use context-free grammars to define language syntax, we will call them ``grammarware engineering tools'' after the paper \cite{Grammarware}. All these tools use grammar definitions and (according to the same paper) there is a strong need in applying software engineering practices in this area. In the present paper we address a problem of modularity and reuse of grammar definitions.

All the grammarware engineering tools have to support reuse of their input artifacts, but it requires tools' authors quite an effort to implement it. We examined popular parser generators \cite{YACC,ANTLR,COCOR,SableCC,xText,SDF,LISA,Rats} and only three of them \cite{SDF,LISA,Rats} have strong reuse capabilities, though even they could be improved in this sense. And grammarware engineering is not limited to parser generators. May be this is natural: when working on a new tool addressing some syntax-related problem (i. e. implementing a new parsing algorithm or a new concept of a pretty-printer) probably one of the last things a developer has on his/her ToDo list is grammar definition reuse, since it is a complicated feature which is mostly irrelevant to what he or she is working on. Anyway it is not likely to appear in the first version.

In UNIX world this problem is solved by the following principle \cite{UNIX}:
{\it Make each program do one thing well.}
Probably it would be ideal if all the grammarware tools could use a common grammar definition language with a common solution for reuse problems. Then it would be easy to support modular and reusable syntax definitions and in addition all the tools would have a common data format by using which they could interoperate with each other. 

To make a step towards this solution we propose a common grammar definition language, named Grammatic, that provides strong modularity and reuse capabilities out of the box. 

One of the main problems in making it suitable for a wide range of tools is that each tool requires different information to be attached to a grammar. Almost no tool takes a mere EBNF definition as input, each one extends it with some extra data. To cope with this Grammatic allows to extend a grammar definition with arbitrary metadata which can be represented in a common format and attached to a grammar externally for the sake of separation of concerns.

An author of a new tool may use Grammatic as follows:
\begin{itemize}
	\item Use Grammatic's grammar definition language to define grammars.
	\item Define extensions for grammar definitions in terms of metadata.
	\item Write custom back end for processing the definition using Grammatic's API.
\end{itemize}
This allows the developer to implement modularity and reuse features easily and concentrate on his or her tool's specific functionality.

But there are many great tools already. They are rarely strong in terms of reuse and even more rarely interoperate well with each other. To benefit both from such tools and Grammatic we can use the latter as a front end, namely we can:  
\begin{itemize}
    \item Identify extensions which the tool adds to a pure grammar definition language.
    \item Decide how to express those extensions with metadata attached to grammar elements.
    \item Write a generator which converts a properly annotated Grammatic grammar definition into the tool's input language.
\end{itemize}
After doing this we can use modular grammar definitions throughout the development process. A not necessarily modularized input for the tool in question is generated only when needed and is never modified by hand.

In this paper we present a case study on the latter case: we demonstrate using Grammatic as a front end for the ANTLR parser generator \cite{ANTLR}. ANTLR is very popular due to its flexibility, clearness and many target languages supported. On the other hand it lacks modularity and supports reuse rather weakly.

The paper is organized as follows: in the next section we give a short overview of Grammatic's main features. Section \ref{Case study} gives an overview of the case study. Subsection \ref{Straightforward} describes a simple way of attaching Grammatic to ANTLR mentioned above and subsection \ref{Sophisticated} describes creating of a more usable though somewhat less general parser generator with Grammatic and ANTLR. Some concluding remarks are given in section \ref{Conclusion}.
\section{Grammatic's features overview}
Here we give an overview of four languages constituting Grammatic's core. These languages are used to define modular grammars and attach metadata to them. 




\subsection{Grammar definitions}
A grammar definition language allows to define a context-free grammar as a set of symbols each of which is associated to a set of productions (in concrete syntax we separate productions by ``$|$$|$''). Here is an example grammar definition which we will use throughout this paper (it describes a simple language of constants and typed variables with assigned values in form of arithmetic expressions):

\begin{verbatim}
    const : ID '=' sum ';' ;
    varDecl : type ID ('=' sum)? ';' ;
    type : ID;
    sum : mult ('+' mult)* ;
    mult : factor ('*' factor)* ;
    factor : NUM || ID || '(' sum ')' ; 
    ALPHA : ['a'--'z' 'A'--'Z' '_'] ;
    ID : ALPHA (ALPHA | ['0'--'9'])* ;
    NUM : ['0'--'9']+ ;
\end{verbatim}
 
In this example characters in single quotes represent embedded lexical definitions. There is no separate notion of a lexical rule (since it is not necessarily required, see \cite{SGLR,Packrat}) and we use the same syntax for EBNF and regular expressions. In this grammar symbol \symbol{sum} is (virtually) nonterminal and \symbol{ID} is (also virtually) terminal since it has only regular expressions on the right side.

\subsection{Imports and Templates}
As we told above, Grammatic's grammar definition language provides strong reuse techniques. Ideas behind these techniques are generalized from the ones implemented in Rats! \cite{Rats}, SDF \cite{SDF} and LISA \cite{LISA}. First we focus on reusing grammar definitions themselves.

The most popular way of reuse is importing. Some grammar definition A might be imported into some other grammar definition B. This means that all the rules of A are inserted into B. Rules of B may refer to symbols of A -- this is the way two grammar definitions are connected. 

Very frequently we have to customize some of the imported rules, i.e. add some more productions to the same symbols or replace existing productions. In paper \cite{ANTLR/Reuse} this is referred to as rule overriding. In Grammatic we decided to use more general form of this concept, namely templates.

A language of grammar templates allows creating grammar definitions with ``placeholders'' which can be replaced with actual objects upon template instantiation. Placeholders might be defined for roles of identifier, expression, production or symbol. A template instantiation might result into grammar object of the type specified by template declaration. Here is an example of a template and its usage. 

\begin{verbatim}
Symbol binaryOperation<ID $name, Expression $sign, Expression $argument> {
    $name --> $argument ($sign $argument)*;
}

import binaryOperation<Product, '*' | '/', Factor>;
import binaryOperation<Sum, '+' | '-', Product>;
Factor
    --> NUMBER
    || ID
    || '(' Sum ')'
    ;
\end{verbatim}

In this example we define a template named ``binaryOperation'' which makes up an infix binary operation out of symbol name, sign and argument expression. Then we instantiate it twice and import instantiation results into current grammar definition -- so we can use new symbol ``Product'' to create ``Sum'' and ``Sum'' to define ``Factor''. 

How can we use templates for ``overriding'' things? We can put a customizable set of rules into a template, provide a placeholder for production or subexpression that should be replaced and then put a right thing in upon instantiation.

\begin{verbatim}
Symbol attributeValue<Production* $moreValueTypes> {
    AttributeValue
        --> STRING
        || ID
        || INT
        || Annotation
        || ValueSequence
        || $moreValueTypes
        ;
}

import attributeValue<
    '{{{' Expression '}}}'
>;
\end{verbatim}

This defines a template for ``AttributeValue'' symbol and instantiates it adding a new production (to use expressions as attribute values).

\subsection{Metadata}
As we told above Grammatic allows to attach arbitrary metadata to a grammar definition in order to express various extensions used by specific tools.

Metadata annotations might be attached to a grammar, symbol, individual production or a subexpression.
Each annotation may contain several attributes (name-value pairs). 
Attribute values may be of different types. There are several predefined value types:
\code{ID}, \code{STRING}, \code{INTEGER}, \code{TUPLE} (a number of name-value pairs) and \code{SEQUENCE} of values and punctuation symbols. 

\begin{verbatim}
  id = someName;       // ID
  str = "some string"; // STRING
  int = 10;            // INTEGER
  class = {            // TUPLE(name : ID; super : ID) 
      name = MyClass; 
      super = Object;
  };          
  astProduction = {{   // SEQUENCE
      ^('+' left ^('-' right 10)) 
  }}; 
\end{verbatim}

Users may add their own types. No attribute itself has any fixed semantics. Metadata is passive, some tools (like analyzers, transformers, generators etc.) may use it according to their needs.

Even without adding custom types many things might be expressed by such annotations. The most powerful type is \code{SEQUENCE} -- it allows to define small embedded DSLs inside Grammatic. We use such DSLs to describe complicated custom properties (see section \ref{Sophisticated}).

\subsection{Queries}

How to attach metadata to a grammar? In many cases it is done by directly embedding annotations into grammar definition. Therefore different concerns are mixed together and this results into a problem: system is not modular, is hard to understand and extend.

We employ ideas from aspect-oriented programming paradigm (AOP, see \cite{AOP}) to solve this problem. In Grammatic a grammar definition itself knows nothing about metadata. All the metadata is attached ``from the outside''. In AOP this is done by defining join points which are described by point cuts \cite{AspectJ}. A language of point cuts is a kind of ``addressing'' notation -- a way to find some object. When we have found such an object we may attach metadata to it (or perform other actions, see below).

In Grammatic we have a language analogous to AOP point cuts -- we call it a query language. For example this query matches rules defining binary operations:

\begin{verbatim}
  #Op --> #Arg (#Sign #Arg)* ;
\end{verbatim}

All the names here represent variables. This query matches any of the following rules:

\begin{verbatim}
  sum : mult ('+' mult)* ;
  mult : factor ('*' factor)* ;
\end{verbatim}

By default a variable matches a symbol but it may match a subexpression or a whole production.

\begin{verbatim}
  Symbol $production:--> $alt:(A | B) ;
\end{verbatim}

Here variables A, B and C match symbol references and Alt matches a subexpression ``B $|$ C''.

We can use wildcards in queries. The following query matches immediately left-recursive rules:

\begin{verbatim}
  #Rec --> #Rec .. ;
\end{verbatim}

Two dots represent a wildcard which matches arbitrary subexpression.

We can consider metadata in our queries. We can restrict a particular attribute to a certain type or value or require attribute's presence or absence:
\begin{verbatim}
  #N {
      type = Nonterminal;
      operation;
      associativity : ID;
      !commutative;
  }
\end{verbatim}

This query matches a symbol with ``type'' attribute having value ``Nonterminal'', ``operation'' attribute present, ``associativity'' attribute having value of type ID and ``commutative'' attribute not present.

\subsection{Aspects}

When a query selects some objects from a grammar definition, we can attach some metadata to them.

\begin{verbatim}
#Rec --> #Rec ..;
[[
    Rec {
        leftRecursive;
    };
]];
\end{verbatim}

This rule adds a ``leftRecursive'' attribute (with no value) to all the symbols matched by Rec variable of this query. A set of such rules constitutes an ``aspect''. Many aspects (independent or not) might be assigned to a single grammar, and even to many grammars since our queries are not tied to concrete objects but only to a grammar structure.

Aspects themselves might be generally reusable -- as we told above, query language does not require ``hard linking'' to grammar objects, these objects are located by their structural context and properties. The rule in our example constitutes a reusable aspect -- we can use it on any grammar.
\section{A Case Study: ANTLR}\label{Case study}
One of the most popular Java-targeted parser generators now is ANTLR \cite{ANTLR}. It is a mature tool based on LL(*) recursive descent parsing algorithm which is empowered by syntactical predicates and backtracking. Many projects, including Sun's NetBeans use ANTLR to generate their parsers. 

On the other hand, ANTLR has some weaknesses in the sense of modularity and reuse. The main issue is that it uses embedded semantic actions, which means that the syntactical structure of the language is physically mixed with Java code describing semantic actions. Thus ANTLR grammars look bloated and grammar structure is not clear. There are also some issues with grammar reuse capabilities though they are being resolved in newer versions (see \cite{ANTLR/Reuse}).

We want to use ANTLR's powerful features but working with modular grammar definitions and having Java code clearly separated from the grammar.

Further sections describe how this could be done with Grammatic.

\subsection{A Straightforward Solution}\label{Straightforward}
A general way of achieving this with Grammatic was described in section \ref{Intro}: we can identify ANTLR's extensions to EBNF, express them in Grammatic's metadata and write a generator to convert annotated Grammatic definitions to ANTLR input language. 

Now let us look at the ANTLR's extensions to EBNF. For the sake of brevity we focus on the most valuable of them here:
\begin{itemize}
    \item Specifying rule parameters and return types.
    \item Embedding semantic actions written in Java.
    \item Specifying syntactic predicates.
\end{itemize}

To express these extensions with metadata we define the following attributes to be used with grammar elements:
\begin{itemize}
    \item Rules: 
        \begin{itemize}
            \item \code{returns : ID;} -- return type.
            \item \code{params : SEQUENCE of TUPLE(type : ID; name : ID);} -- parameters.
        \end{itemize}
    \item Productions:
        \begin{itemize}
            \item \code{predicate : STRING;} -- syntactic predicate for the production.
            \item \code{before : STRING;} -- semantic action to be performed before the production.
            \item \code{after : STRING;} -- semantic action to be performed after the production.
        \end{itemize}
    \item Expressions:
        \begin{itemize}
            \item \code{after : STRING;} -- semantic action to be performed after the expression.
        \end{itemize}
    \item Rule calls:
        \begin{itemize}
            \item \code{arguments : SEQUENCE of ID} -- arguments for the rule call.
        \end{itemize}
\end{itemize}

We define semantic actions as simple strings and it is very close to how ANTLR actually treats them. 

For example let us define an aspect which assigns ANTLR metadata to our arithmetic expressions grammar (see above). We want to have a parser which computes a value of the expression being parsed. Thus our semantic actions will perform arithmetic operations and return values of type \code{int}.
Here is a sample metadata assignment for the rule \symbol{sum}:
\begin{verbatim}
    sum [[returns = int;]]
        $:--> ..
        [[  
            before = '##result = 0;';
            #mult.after = <<
                ##result += #mult;
            >>; 
        ]];
\end{verbatim}
This assigns a \attribute{returns} attribute to the symbol \symbol{sum} itself, \attribute{before} semantic action to the production and \attribute{after} semantic action to all occurrences of \attribute{mult} on the right side. In action bodies we have \code{\#\#result} and \code{\#mult} which correspond to a result variable of a rule and a variable that denotes a value returned by \symbol{mult}. These semantics is to be defined by a generator which will convert our Grammatic definition into ANTLR's language because it depends only on how this generator will treat metadata assigned to grammar elements.

We can handle syntactic predicates the same way. To get the following ANTLR definition:
\begin{verbatim}
NEWLINE
  : ('\r'? '\n')=> '\r'? '\n'
  | '\r'
  ;
\end{verbatim}
We define a grammar rule:
\begin{verbatim}
    NEWLINE : '\r'? '\n' || '\r';
\end{verbatim}
And a metadata assignment rule:
\begin{verbatim}
    NEWLINE
        $:--> ..
        [[
            predicate = <<'\r'? '\n'>>;
        ]]
        --> '\r';
\end{verbatim}
This metadata should also be properly treated by the generator.

Other specific features of ANTLR (like grammar names, Java imports etc.) can be expressed the same way. This method is general enough to be applied in all cases we can imagine and all it adds to the original tool is separation of concerns and reuse techniques available in Grammatic. However, the generator may add some more value to the original tool. We show an example below.

\subsection{A More Sophisticated Solution}\label{Sophisticated}
Nowadays a programming language is usually supported by a strong IDE which makes common activities easier. For example, features brought by the Eclipse IDE for Java \cite{Eclipse} include syntax highlighting, code completion, semantic highlighting, templates, refactorings and many more things. That's why a developer won't be glad to enter Java code outside a specialized editor, say, in ANTLR editor or Grammatic editor. Although these ``non-Java'' editors may provide basic features like highlighting and folding, they are unlikely to provide refactorings and other complicated features. Therefore we want to separate all the Java code from grammar definitions in such a way that it could be edited separately -- in Java editor, using all of its power.

Some tools \cite{SDF,SableCC,ANTLR} solve this problem by generating a parser that builds an AST which is to be processed by external code. This approach has the following disadvantages: it consumes memory for storing AST and time for walking through it. There is also another drawback: many parsers for DSLs simply build models which are very close to ASTs but slightly different (have cross-references, specific additional attributes etc.), in this case a work done by an AST transformer (a program which converts an AST into a model) is simply a waist of resources since all the additional information might be assigned during the parsing process. Thus we want to avoid building such ASTs.

Instead of making a parser always build an AST we propose to use ``Builder'' design pattern \cite{GoF}: a parser should only call methods of some interfaces (builders) which are implemented outside it. Builder interfaces abstract semantic actions of the parser, they are generated along with the parser's code. To illustrate this a bit let us have a look at our \symbol{sum} rule (in ANTLR):
\begin{verbatim}
  sum returns [int result]
    : {result = 0;} 
      left=mult {result += left;} 
      ('+' right=mult {result += right;})*
    ;    
\end{verbatim}
Semantic actions can be abstracted here like this:
\begin{verbatim}
  sum returns [int result]
  @init {
    ISumBuilder builder = myBuilders.getSumBuilder();
  }
    : {builder.init();}
      left=mult {builder.left(left);}
      ('+' right=mult {builder.right(right);})*
      {result = builder.getResult();}
    ;
\end{verbatim}
This is more verbose than immediately embedded actions but this should be generated, no one is to write it by hand.

This approach requires less memory and time since we do not need to build AST objects (which requires memory consumption proportional to input length) and traverse across them. All we need is to create builder objects: this requires us to build only one object for each call that is simultaneously present in the call stack, so it requires memory consumption proportional to the stack depth.

How Grammatic can help us? We are going to define metadata which will give a generator enough information to generate builder interfaces and an ANTLR grammar definition with embedded builder calls.

What metadata do we need to be able to generate builders along with ANTLR grammar? The following information is sufficient:
\begin{itemize}
    \item Return values and parameters for each rule.
    \item Arguments for each rule call.
\end{itemize}

To give a more illustrative example (and create a more flexible system) we will also allow many rules for each grammar symbol. This is useful since we can have only one signature specification (parameters and return value) for each syntactical rule, but the rule might have different semantics when called in different contexts. For example (although a bit strained) we may distinguish constant expressions from ones containing variables, since constant ones may be calculated at compile time. When writing a compiler, we want constant expressions to be evaluated in place (the rule must return a value) and variable expressions are to be stored as objects (expression trees). Hence, from one grammar rule

\begin{verbatim}
    sum : mult ('+' mult)*;
\end{verbatim}
we get two rules with different return types and parameters:
\begin{verbatim}
    varSum [Scope scope] returns [Expression result]
      : varMult[scope] ('+' varMult[scope])* ;

    constSum [Context context] returns [int result]
      : constMult[context] ('+' constMult[context])* ;
\end{verbatim}
(We assume that \code{Scope} maps variable names to objects denoting variables and \code{Context} maps constant names to values.)

We do not want to duplicate rules in our grammar for the sake of these matters, so we will express this in metadata for a single rule. Here is an example of how we can do it:
\begin{verbatim}
sum
    [[
        builders = {{
            Expression varSum(Scope scope);
            int constSum(Context context);
        }};
    ]]
    --> mult ..
    [[
        #mult.call = {
            varSum = {{varMult(scope)}};
            constSum = {{constMult(context)}};
        };
    ]];
\end{verbatim}
What we see here is a small DSL inside Grammatic metadata (actually there are two DSLs: one for specifying return types and parameters and another one for specifying called rules and arguments). The \attribute{builders} attribute of a symbol defines signatures (names, return types and parameters) of ANTLR rules generated for this symbol (two rules will be generated in this example), the \attribute{call} attribute of a symbol reference  specifies an ANTLR rule (with arguments) which should be called in each case. 

The brevity of metadata definition given above is achieved through Grammatic's ability to define internal DSLs. This is done by parsing attribute values of type \code{SEQUENCE} by an externally supplied parser. Lexical structure of these DSLs is fixed: sequence elements (identifiers, strings, numbers, tuples, sequences and punctuation values) serve as tokens.

A generator will produce two rules given above form this example. Above we omitted builder calls from rule definitions to be more clear. A full rule looks like this:
\begin{verbatim}
    varSum [Scope scope] returns [Expression result]
    @init {
        IVarSumBuilder builder = myBuilders.getVarSumBuilder(scope);
    }
      : vm=varMult[scope] {builder.varMult(vm);} 
        ('+' vm1=varMult[scope] {builder.varMult(vm1)})* ;      
\end{verbatim}


\section{Conclusion and Future Work}\label{Conclusion}

This paper addressed solving problems of modularity and reuse of grammar definitions by defining a general front end, Grammatic. This front end can be adopted by newly developed tools through its API or it can be attached to an existing tool by creating a converter from its universal format to the tool's specific input format. Grammatic provides a language for defining modular grammars (supporting templates and imports), which is extensible by attaching arbitrary metadata. It also supports separation of concerns by defining reusable aspects.

We showed two ways of using Grammatic to bring reuse and separation of concerns to a popular parser generator -- ANTLR. The most straightforward way is based on expressing extensions added by the tool to a general grammar definition language in terms of metadata and creating a generator which transforms an annotated Grammatic's definition into the tool's input. We also presented a way of using Grammatic to separate custom code written in the target language (Java) from the grammar definition and other metadata. This is done by using ``Builder'' design pattern: generating a set of interfaces which are to be implemented manually. This allows a developer to use the power of his IDE when working with Java code.

The case study shows that Grammatic helps adding reuse and modularity capabilities to existing tools. We plan to apply such practices to some more tools to find out more things to be supported by Grammatic. For now we plan to support metadata templates, grammar testing facilities, text generation and error tracking facilities helping to convert errors reported by a back end to Grammatic's errors.

Our long term goal is to create a common grammar definition platform usable by a wide range of grammarware engineering tools.
\bibliographystyle{unsrt}
\bibliography{grammatic_main}

\end{document}